\documentclass[]{spie}  

\usepackage{amsmath,amsfonts,amssymb}
\usepackage{graphicx}
\usepackage[colorlinks=true, allcolors=blue]{hyperref}

\usepackage{color}

\title{A fast method for the detection of vascular structure in images, based on the continuous wavelet transform with the Morlet wavelet having a low central frequency}

\author[a]{Eugene B. Postnikov}
\author[b]{Maria O. Tsoy}
\author[b]{Maxim A. Kurochkin}
\author[b]{Dmitry E. Postnov}
\affil[a]{Kursk State University, Department of Theoretical Physics, Radishcheva st. 33, Kursk 305000, Russia}
\affil[b]{Department of Optics and Biophotonics, Saratov State National Research University, Astrakhanskaya st. 83, Saratov 410012, Russia}
\authorinfo{Corresponding author: E.B. Postnikov. E-mail: postnicov@gmail.com}

\begin{document} 
\maketitle

\begin{abstract}
A manual measurement of blood vessels diameter is a conventional component of routine visual assessment of microcirculation, say, during optical capillaroscopy. However,  many modern optical methods for blood flow measurements demand the  reliable procedure for a fully automated detection of vessels and estimation of their diameter that is  a challenging task. 
Specifically, if one measure  the velocity of red blood cells by means of laser speckle imaging, then  visual measurements become   impossible, while the velocity-based estimation has their own limitations.   One of promising approaches is based on fast switching of illumination type, but it drastically reduces the observation time, and hence, the achievable quality of images.   
In the present work we address this problem proposing an alternative method for the processing of noisy images of vascular structure, which extracts the mask denoting  locations of  vessels, based on the application of the continuous wavelet transform with the Morlet wavelet having small central frequencies. Such a method combines a  reasonable accuracy with the possibility of fast direct implementation to images. Discussing the  latter, we describe in details a new MATLAB program code realization for the CWT with the Morlet wavelet, which does not use loops completely replaced with element-by-element operations that drastically reduces the computation time.
\end{abstract}


\section{Introduction}
 
 Being the conventional component of some routine methods of  medical screening, say, during capillaroscopy \cite{Maricq1973},  a measurement of blood vessels diameter  becomes  the key technique for everyone who needs to quantify the  functioning of vascular networks.  While laser-speckle imaging (LSI) generally shows only the relative changes of red blood cells (RBC) velocity \cite{Tuchin2013},
some other methods, like particle image visualisation (PIV) can deliver its  absolute values. The computational scheme of both these methods implies the selection of regions of interest (ROI), to avoid the useless and computationally expensive processing of ``empty'' image segments. Here the segmentation of the micrivascular network together with the fast and reliable determination of margins of vessels in images is highly demanded.   
While there are the few published attempts to  estimate stationary vessel diameter from laser speckle data  \cite{Liu2014}, 
 the problem becomes even more complicated if one needs to monitor the  microcirculatory network dynamics, say in response to different stimuli \cite{Neganova2016}. 
One of the promising approaches is based on the fast switching of illumination type, but it drastically reduces the observation time, and hence, the quality of images. For LSI applications, recent attempt is known to avoid such problems \cite{Postnov2016}.
 
The wavelet transform, which is a  powerful tool for the signal and image processing being especially adjusted to the detection of multiscale patterns \cite{MallatBook}, attracted an attention in biomedical image processing even at first stages of its development \cite{Aldroubi1996book}. Now the application of wavelet methods on parity with other image filtering approaches is widespread and included into standard mathematical software as MATLAB \cite{Semmlow2014book}. 

Being initially aimed for image compressing, these wavelet-based  methods are focused mainly on the discrete wavelet transform, which inherits classical filtering algorithms. In addition, there usually operate in the domain of real functions that gives multiscale expansions, which require an additional processing to reveal localized structures, see the reviews in books cited above. Some known attempts to utilize the discrete complex wavelet transform, which follow the pioneering work \cite{Kingsbury1999}, meet certain difficulties operating with signals without the binary hierarchical subdivision. 
At the same time, the complex continuous wavelet transform is more flexible in the sense of arbitrary continual spatial shifts and scales but conventionally used mainly for the detection of periodic pattern but not to identify the individual spots and curves. 

In our work, we explore how the spatial-domain continuous wavelet transform  can be used on the playground of the pattern recognition typical for the processing of images of microcirculatory network. Specifically, we apply it for the automated determination of a ``recommended-to-process'' part of both the surrogate spatial pattern and for the  real vascular network image obtained from the chorioallantoic membrane (CAM) of chicken embryo. The obtained results show the applicability of suggested approach and suggest the way for further its development.

\section{Method}

\subsection{Premises of the approach}

The integral representing the continuous wavelet transform (CWT),
\begin{equation}
w(a,b)=\int_{-\infty}^{+\infty}f(t)\psi^*\left(\frac{t-b}{a}\right)dt,
\label{CWT}
\end{equation}
has a direct mathematical interpretation as a correlation between the analysed function $f(t)$ and the wavelet $\psi$ (the asterisk in (\ref{CWT}) denotes a complex conjugation) calculated for the given shift $b$ and scale $a$. Thus, its maxima correspond to shapes of the function's features, which locally resemble the wavelet's shape in a best way. 

\begin{figure}%
\includegraphics[width=\columnwidth]{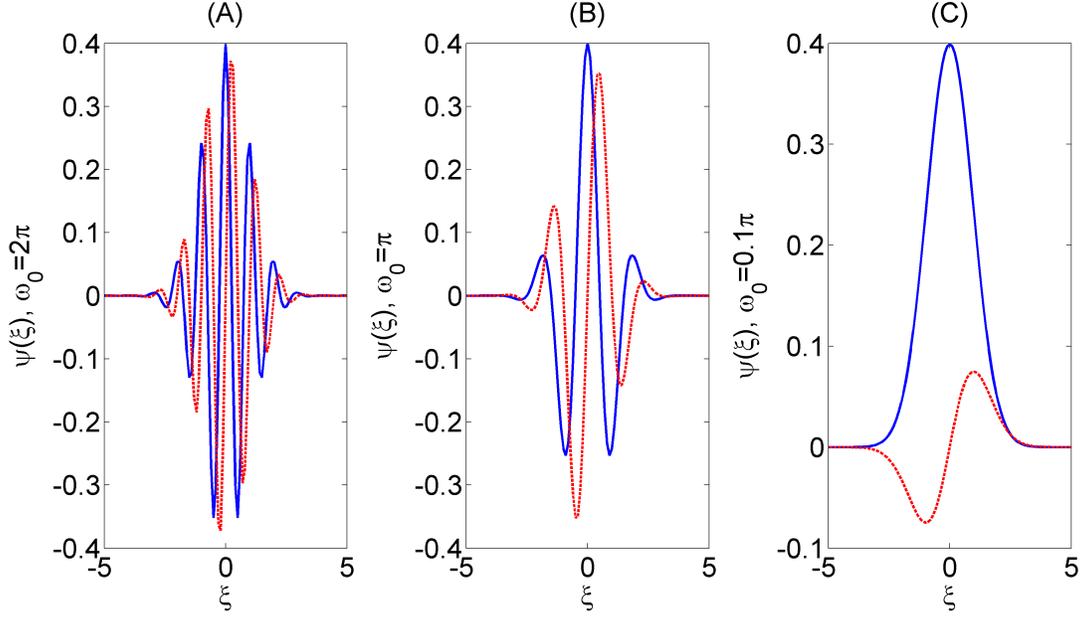}%
\caption{The Morlet wavelet for different central frequencies $\omega_0$. The real and imaginary parts are shown as solid and dashed lines, respectively.}%
\label{figpsi}%
\end{figure}

Among a variety of wavelet functions, the standard Morlet wavelet defined in the amplitude norm as 
\begin{equation}
\psi(\xi)=\frac{1}{\sqrt{2\pi}}e^{i\omega_0\xi}e^{-\frac{\xi^2}{2}}
\label{defMorlet}
\end{equation}
is one of the most popular for the local spectral analysis, i.e. for revealing local periodicities. Such an application is based on the shape of this wavelet (\ref{defMorlet}), which is the harmonic function modulated by the Gaussian. It exhibits a number of oscillations for the conventionally used central frequencies $\omega_0\geq 2\pi$, see Fig.~\ref{figpsi}(A). These large values are based on the so-called admissibility condition $|\widehat{\psi(0)}|=0$, which is satisfied for (\ref{defMorlet}) only approximately: $|\widehat{\psi(0)}|=\exp(-\omega_0^2/2)$.

On the other hand, it has been shown recently \cite{Lebedeva2014,Postnikov2016} that the admissibility condition is not a necessary property, and its violation does not prevent an exact invertibility of the transform. Therefore, we can feel free to consider smaller values of the central frequency, which result in low- and non-oscillating shapes of the wavelet (\ref{defMorlet}). 

I should be pointed out that the complex conjugated expression (\ref{defMorlet}) can be expanded in the Tailor series and truncated for small $\omega_0$ up to the first terms as
\begin{equation}
\psi(\xi)=\frac{1}{\sqrt{2\pi}}e^{-\frac{\xi^2}{2}}-
i\frac{\omega_0}{\sqrt{2\pi}}\xi e^{-\frac{\xi^2}{2}}=
\left[1+i\omega_0\frac{d}{d\xi}\right]\frac{1}{\sqrt{2\pi}}e^{-\frac{\xi^2}{2}}.
\label{expMorlet}
\end{equation}
One can see that the real part of (\ref{expMorlet}) being substituted into Eq.~(\ref{CWT}) provides a Gaussian (diffusion) smoothing of the processed function $f(t)$, i.e. its de-noising; and the imaginary part evaluates differentiation of the smoothed signal, i.e. edge detection. See Fig.~\ref{figpsi}(C) for a typical example of the shape of these sliding windows. 

At the same time, the basic property of the CWT with the Morlet wavelet applied to the pure harmonic function, is the localisation of the signal's frequency as a maximum of the Gaussian representing CWT's modulus with respect to the scale:
$$
\left|w(a,b)\left[C\exp(i\omega t)\right]\right|=Ce^{-\frac{(a\omega-\omega_0)^2}{2}}.
$$ 
This property is fulfilled for arbitrary $\omega_0$. Note also that the series of wavelet coefficients corresponding to $a_{max}=\omega_0/\omega$ exactly coincides with a magnitude of the analysed signal $\left|w(a_{max},b)\right|$ as well as a phase factor of the CWT. Thus, the determination of $a_{max}$ automatically provides values of the desired features too. 

Therefore, the CWT with small central frequencies simultaneously de-noises localized maxima of the signal bounded by some edges and allows for determining their scales simply detecting CWT modulus maxima.   

\subsection{Algorithm of image processing}

The proposed algorithm is formulated as follows:
\begin{itemize}
	\item determine number $N$ of pixels in the image’s row, which give the scaling of frequencies of the Fourier transform $\omega_k=2\pi[0..\mathrm{ceil}(N/2)]/(N-1)$ (here we use the positive frequencies only, whence  $\mathrm{ceil}(N/2)$ rounds $N/2$ towards plus infinity);
	\item to fix a small central frequency, e.g. $\omega_0\leq \pi$ and an appropriate matrices of the proper scales $a_j$ and Gaussian factors $g_{jk}=\exp\left[-(a_j\omega_k-\omega_0)^2)/2\right]$;
	\item for each row
	\subitem i) to find the Fourier coefficients $F(\omega_k)$ of the deviations of pixels intensities from its average value in a row via the Fast Fourier Transform (FFT); compute their elementwise Gaussian filtering 
	$F(\omega_k)g_{jk}$, and the inverse Fourier transform of the results doubled to take into account only half of frequencies used; this results in a matrix of the CWT coefficients $w(a_j,t_n)$, $n=1..N$.
	\subitem ii) to find a set of maxima $\{a_{jmax}\}$ for each array $w(a_j,t_n)$ with a fixed $n$ and to reconstruct the envelopes of vasa's cross section pictures by the summation 
	$\sum_{jmax} w\left(\{a_{jmax}\},t_n\right)$ and taking of real part of the result;
	\item the obtained image should be binarised with respect zero (or small threshold above zero) level; the positive values will be mapped into 1's, which provide the desired mask identifying vessels locations.    
\end{itemize}

For better accuracy, the procedure can be repeated along columns of digital images (in this case vasa situated more horizontally will be detected; they may be missed during the row-by-row processing) and both processed images combined to obtain a more accurate mask. 

\subsection{Speeding up CWT computation with MATLAB} 

It should be pointed out that the algorithm described above includes a number of string-by-string and row-by-row loops being applied to 2D images processing. This fact naturally induces  its practical realization using software utilizing internal matrix-based methods, which provides loopless fast computations, fro example MATLAB can be considered in such role. 

We propose new realization of the CWT with the Morlet wavelet, which replaces \verb!for!-loop-based algorithms utilized in the function included into the conventional packages such as {\tt Wavelet Toolbox} by MathWorks and {\tt WaveLab} by the Stanford University. 

Then the program code (and comments marked by \% sign, which describe all variables) reads as follows:
\begin{verbatim}
% Input parameters:
% u    -    input data (a row array)
% N    -    length of u
% omega0  - the central frequency's value
% a    -    scale values 
% Na -      length of a
% Output parameter:
% w  -    the resulting complex-values matrix of the wavelet trasform of the size N*Na
L=ceil(N/2);%1
omega_=2*pi*(0:L)/(N-1);%2
AA=repmat(a',1,L+1);%3
OMEGA2=repmat(omega_,Na,1);%4
OMEGAs2=OMEGA2.*AA;%5
WINDOW2=exp(-(OMEGAs2-omega0).^2/2);%6
F=fft(u);%7
FF2=repmat(2*F(1:L+1),Na,1);%8
CNV2=WINDOW2.*FF2;%9
w=ifft(CNV2,N,2);%10
\end{verbatim}

The line \verb!%2! 
form a row array of frequencies, for which \verb!fft! function gives the Fourier coefficients; it is converted into 2D matrix, which contains \verb!L+1! rows and \verb!Na! columns in the line \verb!%4! 
by repetition of columns containing frequency values. The same procedure generates 2D matrix of scales of the same shape (the line \verb!%3!).   
As a result, the set of scaled frequencies and windows (the lines \verb!%5! and \verb!%6!) 
are formed by the element-by-element multiplication (marked as .*) instead of any time-consuming loop. This part of code is the same for all rows of an image, therefore, can be calculated one time and placed in the beginning of the full program.

The second part (the lines \verb!%7!--\verb!%10!) 
directly calculates the CWT using the predefined common parameters. These lines of code contain the Fourier transform via FFT algorithm (\verb!%7!), 
forming 2D matrix corresponding to its positive frequencies with the number of columns equal to the number of used scales (\verb!%8!), 
sliding Gaussian filtering of the spectrum (\verb!%10!)
realized as an element-by-element multiplication, and, finally, the inverse Fourier transform. Note that we kept two times reduced size of all matrices up to this moment. However, the special feature of MATLAB's \verb!ifft! function provides an opportunity not only to invert the the whole matrix by columns but also make them padded with zeros up to the required number \verb!N!.

\section{Results and their discussion}

\begin{figure}%
\includegraphics[width=\columnwidth]{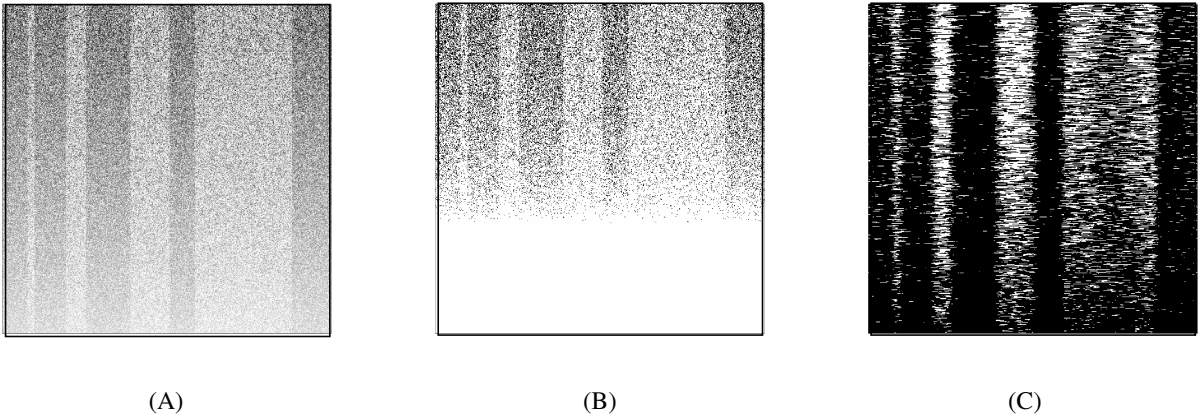}%
\caption{The surrogate noisy image of $512\times 512$ pixels (a) and the results of its processing : 
(b) the  result of the direct (threshold) conversion to a black-and-white image; (c) the re-composed from the wavelet transform  binary image  showing the location of stripes in original image.}%
\label{figtest0}%
\end{figure}

For the testing purposes we first created the surrogate image being composed of (i) a random pattern, (ii) stripes of different width with reduced brightness, and (iii) an intensity gradient applied from top to bottom.  Both the surrogate image and the processing results are illustrated in Fig.~\ref{figtest0}. Obviously, the simple threshold-based method to extract the stripes location and shape, which is shown in panel (b),   fails due to the presence of gradient of overall brightness, which leads to the complete loss of a bottom part of the image. The wavelet-based technique (c) shows much better result in this respect. Note, we intentionally did not provided any smoothing or interpolation to reduce the hatched structure which is visible  of the image in panel (c) in order to show raw result of the transform. Such smoothing  can be done during the subsequent composition with results of vertical scan.  

\begin{figure}%
\includegraphics[width=\columnwidth]{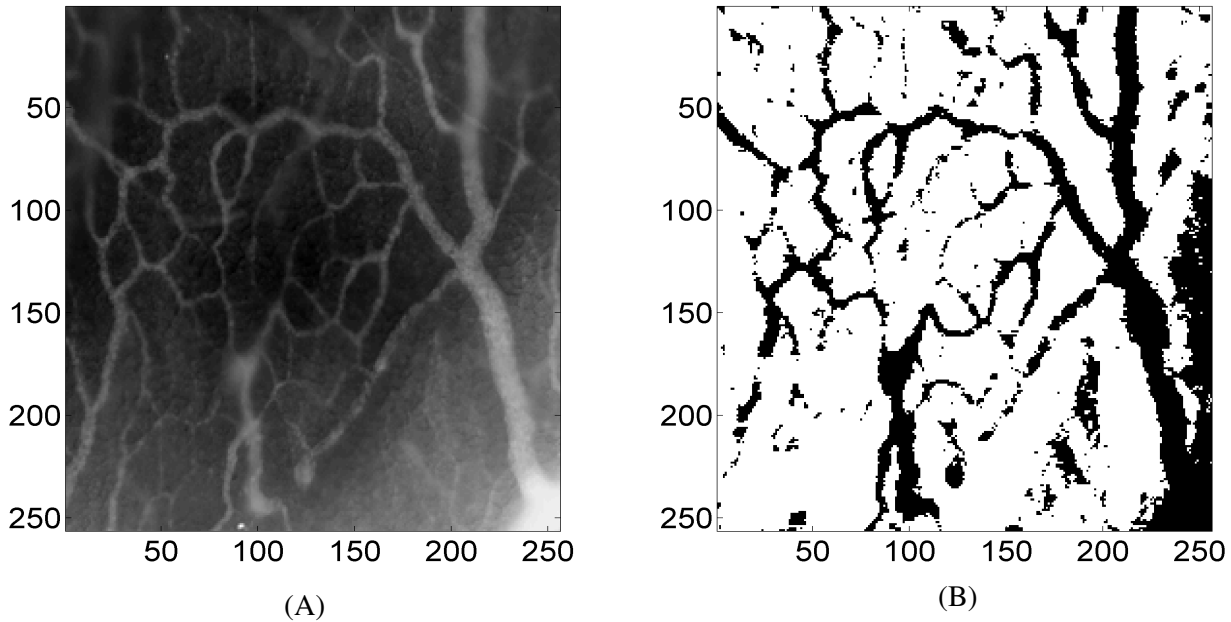}%
\caption{The image of the vascular structure  , $256\times 256$ pixels, (A) and the resulting mask), denoting vessels locations after an evaluation of the proposed CWT-based algorithm along rows (B). Here, black and white pixels mark presence and absence of vessels detected.}%
\label{figtest}%
\end{figure}

In order to apply the method to the real microcirculatory network, we use the image obtained from CAM of 11-day old chicken embryo. The preparation procedure is quite simple and can be found everywhere. The opened egg shell was fixed vertically and illuminated with green light. 
The optical image was recorded in gray scale. The arbitrary chosen single frame is shown in Fig.~\ref{figtest}(left panel). It was processed with the algorithm described above with the central frequency $\omega_0=\pi$, see the corresponding wavelet shape in Fig.~\ref{figpsi}(B). Although this value does not correspond to a single bell-shaped curve like Fig.~\ref{figpsi}(C), the practically one full oscillation presented there provides an effective way to detect a localised bright regions obtained by the cross section of the image. A series of numerical experiments with different $\omega_0$ have demonstrated an effectiveness of such a choice for the present ranges of vessels diameters and spatial noise scales. It can be explained if one note that the chosen shape is quite close to the classical Laplacian filter \cite{Semmlow2014book} and its derivative. Thus, there is actually a kind of combination of Laplacian de-noising and the zero-crossing edge detection algorithm realized simultaneously since the wavelet is 
a complex function.

Fig.~\ref{figtest}(right panel) shows the obtained mask (binarized image). One can see the different degrees of visualization depending on vessel size. This is a sign of the promising method features, not yet fully implemented. Also note,  that not all the vessels are mapped in proportionally reasonable sizes. Definitely, all this is dependent on the vessel orientation and, thus, the corresponding method development is in order.

\section{Conclusion and outlooks}

In conclusion, we have shown a new promising way  for the fast and computationally effective processing of image data obtained from the microscopy of microcirculatory network of chicken embrio. Esentially the same approach can be used in the case of processing not just optical images  but some calculated distributions of quantities relevant to the measured microcirculation dymanics. We believe that the capability of fast and reliable processing will be demanded in many applications. Specificaly, we intend the further use of the described technique for the purposes of pattern separation during the estivation of dynamic changes in vessel barrier functions. While in specific vascular networks like brain or retina,  vessels show high protections from penetration of different chemicals, in available biomodels such leakage can be considerable, and, thus, more easy to be quantified, but at the same time need more precize determination of vessels locations and boundaries. The proposed technique is expected to show 
 good results for this task. 

\section*{Acknowledgements}

This work was funded by Russian Science Foundation, project 16-15-10252.


\begin{thebibliography}{10}

\bibitem{Maricq1973}
Maricq, H.~R. and Carwile~LeRoy, E., ``Patterns of finger capillary
  abnormalities in connective tissue disease by “wide-field” microscopy,''
  {\em Arthritis and Rheumatism}~{\bf 16}(5),  619--628 (1973).

\bibitem{Tuchin2013}
Tuchin, V.~V., ed.,  [{\em Handbook of Coherent-domain Optical Methods:
  Biomedical Diagnosis, Environmental Monitoring, and Materials
  Science}{\nolinebreak\hspace{0.1em}]}, Springer (2013).

\bibitem{Liu2014}
Liu, Q., Li, Y., Lu, H., and Tong, S., ``Real-time high resolution laser
  speckle imaging of cerebral vascular changes in a rodent photothrombosis
  model,'' {\em Biomed. Opt. Express}~{\bf 5},  1483–1493 (2014).

\bibitem{Neganova2016}
Neganova, A., Postnov, D., Brings-Jacobsen, J., and Sosnovtseva, O., ``Laser
  speckle analysis of retinal vascular dynamics,'' {\em Biomed. Opt.
  Express}~{\bf 7},  1375--1384 (2016).

\bibitem{Postnov2016}
Postnov, D.~D., Tuchin, V.~V., and Sosnovtseva, O., ``Estimation of vessel
  diameter and blood flow dynamics from laser speckle images,'' {\em Biomedical
  Optics Express}~{\bf 7},  2759--2768 (2016).

\bibitem{MallatBook}
Mallat, S.,  [{\em A wavelet tour of signal
  processing}{\nolinebreak\hspace{0.1em}]}, Academic press (1999).

\bibitem{Aldroubi1996book}
Aldroubi, A. and Unser, M., eds.,  [{\em Wavelets in medicine and
  biology}{\nolinebreak\hspace{0.1em}]}, CRC press (1996).

\bibitem{Semmlow2014book}
Semmlow, J.~L. and Griffel, B.,  [{\em {Biosignal and Medical Image
  Processing}}{\nolinebreak\hspace{0.1em}]}, CRC Press (2014).

\bibitem{Kingsbury1999}
Kingsbury, N., ``Image processing with complex wavelets,'' {\em Philos. Trans.
  Royal Soc. A: Math., Phys. Eng. Sci.}~{\bf 357}(1760),  2543--2560 (1999).

\bibitem{Lebedeva2014}
Lebedeva, E.~A. and Postnikov, E.~B., ``On alternative wavelet reconstruction
  formula: a case study of approximate wavelets,'' {\em Royal Soc. Open
  Sci.}~{\bf 1},  140124 (2014).

\bibitem{Postnikov2016}
Postnikov, E.~B., Lebedeva, E.~A., and Lavrova, A.~I., ``Computational
  implementation of the inverse continuous wavelet transform without a
  requirement of the admissibility condition,'' {\em Appl. Math. Comput.}~{\bf
  282},  128--136 (2016).

\end{thebibliography}
\end{document}